\documentclass[natbib]{svjour3}                     
\smartqed  
\usepackage{graphicx}
\newcommand{\lesssim}{\mathrel{\hbox{\rlap{\hbox{\lower4pt\hbox{$\sim$}}}\hbox{$<$}}}}
\newcommand{\gtrsim}{\mathrel{\hbox{\rlap{\hbox{\lower4pt\hbox{$\sim$}}}\hbox{$>$}}}}
\begin{document}

\title{General overview of black hole accretion theory}


\author{Omer Blaes}


\institute{O. Blaes \at
              Department of Physics, University of California, Santa Barbara, CA 93106, USA\\
              Tel.: +1-805-8937239\\
              \email{blaes@physics.ucsb.edu}           
}

\date{Received: date / Accepted: date}

\maketitle

\begin{abstract}
I provide a broad overview of the basic theoretical paradigms of black hole
accretion flows.  Models that make contact with observations continue to be
mostly based on the four decade old alpha stress prescription of \citet{sha73},
and I discuss the properties of both radiatively efficient and inefficient
models, including their local properties, their expected stability to
secular perturbations, and how they might be tied together in global
flow geometries.  The alpha stress is a prescription for turbulence, for
which the only existing plausible candidate is that which develops from the
magnetorotational instability (MRI).  I therefore also review what is currently
known about the local properties of such turbulence, and the physical issues
that have been elucidated and that remain uncertain that are relevant for the
various alpha-based black hole accretion flow models.
\end{abstract}

\section{Introduction}
\label{sec:intro}

Accretion is the very process that allows black hole sources to emit
electromagnetic radiation and other forms of energy.  Because black holes
are so small in size compared to the spatial scale of their sources of fueling,
and because centrifugal forces on matter of given angular momentum increase
more rapidly ($\propto R^{-3}$) than gravity ($\propto R^{-2}$) as one
moves inward in radius $R$, accretion is generally believed to be a process
involving rotationally supported flows.  Matter in such a flow must lose
angular momentum in order to move inward and release gravitational binding
energy.  It is the nature of the angular momentum loss mechanism, and the
process whereby gravitational binding energy is converted into observable
forms of energy, that are the two central questions of black hole accretion
theory.  At least three mechanisms have been proposed for angular momentum
extraction:

\paragraph{{\rm (1)} External stresses associated with large scale magnetic
fields in a magnetohydrodynamic (MHD) outflow.}  This mechanism \citep{bla82}
may be
relevant in low luminosity sources where accretion power may be largely
converted into mechanical power in outflows.  It may also be relevant in
resolving the fueling and self-gravity problems in the outer accretion flows
in active galactic nuclei \citep{goo03}.  Whether and how large scale magnetic
fields can be created remains an open question, however.

\paragraph{{\rm (2)} Magnetorotational (MRI) turbulence.}  Such turbulence is
generic for plasmas that are sufficiently electrically conducting
and not too strongly magnetized
\citep{bal91, haw91, bal92, bal98}.  Because turbulence is inherently
dissipative, this process is almost certainly relevant for sources whose
power output is dominated by thermal radiative emission mechanisms.

\paragraph{{\rm (3)} Nonaxisymmetric waves and shocks.}  Nonaxisymmetric
(e.g. spiral) waves can transport angular momentum outward through the
flow.  Such waves can also transport energy away from the region where
gravitational binding energy is released, depositing it elsewhere.
Waves are almost certainly relevant in disks around supermassive
black hole binaries, and also in the outer, self-gravitating parts of
disks in active galactic nuclei.  They probably also play a role in the
outer parts of black hole X-ray binary disks due to tidal excitation by
the companion star.  Nonaxisymmetric shocks can also play an important role
in the inner regions of accretion flows whose angular momenta are misaligned
with the black hole spin axis \citep{fra08}.

\paragraph{}

Among these options, only the second - MRI turbulence - is a mechanism
that {\it might} be describable by the classical alpha prescription of
\citet{sha73}, at least in some aspects \citep{bal99}.  The angular momentum
transporting stress $w_{R\phi}$ in the turbulence is given by local space and
time averages of correlated fluctuations in radial ($R$) and azimuthal ($\phi$)
fluctuations of velocity ${\bf v}$ (the Reynolds stress) and
magnetic field ${\bf B}$ (the Maxwell stress),
\begin{equation}
w_{R\phi}=\left<\rho v_R\delta v_\phi-{B_RB_\phi\over4\pi}\right>,
\end{equation}
where $\rho$ here is the mass density and $\delta v_\phi$ is the local
deviation of the azimuthal velocity component
from the mean background shear flow.  Maxwell stresses are generally larger
in magnitude than the Reynolds stresses by factors of at least several.
I say that the total stress {\it might} be describable by the
classical alpha prescription because these stresses appear to be mostly local
in the sense that simulations show that radial correlations in stress drop
rapidly on scales larger than the local disk scale height.  However, as I
discuss in section 3.1 below, there remain correlations on larger radial scales.

This article provides a broad overview of alpha-based models of black hole
accretion flows, focusing on structure, dynamics, and thermodynamics.  These
models continue to dominate theoretical efforts to explain observations, but
a slow revolution is occurring as simulations of
MRI turbulence, both local and global, continue to become more powerful and
to incorporate more and more of the relevant physics.  This article will also
discuss what has been learned recently from local, shearing box simulations of
MRI turbulence as this pertains directly to some of the alpha-based modeling.
A review of global simulations can be found in Chapter 2.4.
Spectral modeling of accretion flows is discussed in Chapter 2.3.
I will also mainly focus
on {\it accretion} rather than the formation of jets and outflows here,
though jets and outflows are clearly important (both observationally and
theoretically, in certain flow states).  See Chapter 5.3 on jet
launching mechanisms.

\section{Hydrodynamic disk models with the alpha prescription}
\label{sec:alpha}

\begin{figure}
\includegraphics[width=1.0\textwidth]{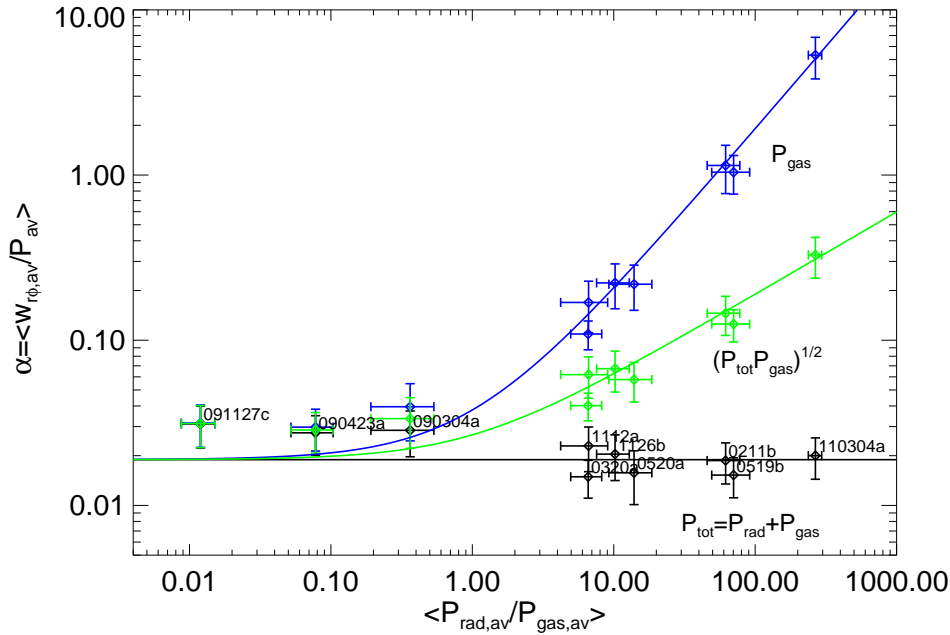}
\caption{Time-averaged values of the ratio of spatially averaged stress
to various measures of spatially averaged thermal pressure, i.e. the
\citet{sha73} parameter alpha, as a function of the time-averaged ratio
of spatially averaged radiation pressure to spatially averaged gas pressure,
for a number of radiation MHD, vertically stratified, shearing box simulations
of MRI turbulence. Black, green and blue points are the results for thermal
pressures defined to be the total (radiation plus gas) pressure, the geometric
mean of the total and gas pressures, and the gas pressure alone, respectively.
Horizontal and vertical error bars on all points indicate one standard
deviation in the respective time-averages.
The horizontal black line is the average alpha value of the total pressure
prescription (black) points, while the green and blue curves are what would
result if the total pressure prescription were correct, but one nevertheless
insisted on defining alpha in terms of the other thermal pressure definitions
used in the green and blue points, respectively.  The stress prescription that
is most consistent with the simulation data is one in which the total thermal
pressure is used, though it is perhaps noteworthy that the alpha values in
the gas pressure dominated simulations are consistently higher than the
alpha values in the radiation pressure dominated simulations.
(Updated from \citealt{hir09a}.)} 
\label{fig:alphavspradopgas}
\end{figure}

Decades of theory and models of black hole accretion flows have critically
relied on the alpha prescription for a local stress introduced by \cite{sha73}.
There are numerous variants of this prescription which produce order unity
changes in the definition of $\alpha$, and one of the most common is
\begin{equation}
w_{R\phi}=\alpha P,
\label{eq:wrphialpha}
\end{equation}
where $P$ is the thermal pressure.  Most models have assumed that this is
the {\it total} thermal pressure (gas plus radiation), but prescriptions in
which the stress is taken to be proportional to just the gas pressure alone
(e.g. \citealt{sak81}) or the geometric mean of the gas and total thermal
pressures (e.g. \citealt{taa84}) have also been suggested.  However,
as illustrated in Figure~\ref{fig:alphavspradopgas}, recent
radiation MHD simulations of MRI turbulence find that the stress scales
best with total thermal pressure, at least on long time scales
\citep{ohs09,hir09a}.

The alpha prescription (\ref{eq:wrphialpha}) is usually
used to solve for the radial structure of vertically-integrated geometrically
thin or slim accretion disks, in which case it enters the equations through
the vertically-integrated stress:
\begin{equation}
W_{R\phi}=\int_{-\infty}^\infty w_{R\phi}dz\sim2H\alpha P,
\end{equation}
where $P$ is now some vertically averaged thermal pressure, of order the
midplane pressure, and $H$ is the vertical half-thickness of the disk.
This is consistent with MRI simulations, but
the prescription is also occasionally used even more locally by assuming that
the {\it vertical} profiles of stress and dissipation at a given radius are
proportional to the local vertical profile of thermal pressure.  As we
discuss below in section 3.2, this is not consistent with vertically
stratified simulations of MRI turbulence, which generally have vertical
profiles of stress that are broader than the thermal pressure profile.
Alpha defined locally
would therefore increase rapidly outward from the disk midplane.

\subsection{Local thermal equilibria and secular instabilities}
\label{sec:localthermalequil}

Virtually all (non-simulation-based)
models of black hole accretion flows are based on vertically
integrated hydrodynamic equations.  These models often
neglect the possibility of significant losses of mass, angular momentum, and
energy in outflows and jets, though some models do attempt to include them
with various prescriptions, particularly in advection-dominated flows which
we will come to shortly.  As discussed in Chapter 5.1,
neglect of outflows is likely to be a bad approximation in some
sources and accretion states. Nevertheless,
if we adopt this assumption for simplicity, then for stationary flows, the
conservation laws of mass, radial momentum, angular momentum, and internal
energy can be written as
\begin{equation}
\dot{M}=2\pi R\Sigma v,
\end{equation}
\begin{equation}
\rho v{dv\over dR}=\rho(\Omega^2-\Omega_{\rm K}^2)R-{dP\over dR},
\end{equation}
\begin{equation}
\dot M{d\ell\over dR}={d\over dR}(2\pi R^2W_{R\phi}),
\end{equation}
and
\begin{equation}
Q_{\rm adv}\equiv{-\dot{M}\over4\pi R}\left[{dU\over dR}+P{d\over dR}
\left({1\over\rho}\right)\right]=Q^+-Q^-,
\end{equation}
where we have neglected general relativity for the purposes of physical
transparency.  Here $\rho\sim\Sigma/(2H)$ is a vertically-averaged
density, $\Sigma$ is the surface mass density, $\Omega$ is the fluid angular
velocity which may differ from the test particle (Kepler) angular velocity
$\Omega_{\rm K}$, $\ell=\Omega R^2$ is the fluid specific angular momentum,
$v$ is the inward radial drift speed, $U$ is a vertical average of the internal
energy per unit mass, $Q^-$ is the radiative cooling rate per unit surface
area on each face of the disk, $Q^+=-(1/2)W_{R\phi}Rd\Omega/dR$ is half the
turbulent dissipation rate per unit surface area, and $Q_{\rm adv}$ is half the
inward radial advection of heat per unit surface area.  Vertical hydrostatic
equilibrium implies that the vertical half-thickness of the disk is
$H\sim(P/\rho)^{1/2}/\Omega_{\rm K}$.

Once one adopts the alpha prescription (\ref{eq:wrphialpha}), together with
an equation of state and opacities and/or optically thin cooling functions,
it is possible to solve these equations with assumed
boundary conditions to derive the radial profiles of vertically-averaged
fluid variables in the flow.  Such models generally invoke a regularity
condition at an inner sonic point and/or a no-torque inner boundary condition
at, for example, the innermost stable circular orbit, although
magnetohydrodynamic stresses can be important once one enters the plunging
region near the black hole \citep{gam99,kro99}.  (See Chapter 2.4 and, e.g.,
\citealt{pen10} and \citealt{nob10} for recent simulation work on this issue for
geometrically thin disks.)  Another approach is to consider radii
much larger than the gravitational radius $R_{\rm g}\equiv GM/c^2$ and
invoke self-similarity by
assuming a constant ratio of advective cooling over turbulent dissipation
$Q_{\rm adv}/Q^+$ \citep{nar94}.

For a fixed black hole mass, models that are stationary generally depend on
a number of chosen parameters, the most important being the accretion rate
$\dot{M}$ which is everywhere constant through the flow (remember, we are
neglecting outflows here).  A number of possible equilibria have been
discovered in this way, and the primary method of choosing which ones are
physically realizable in nature has
been to check if they are stable to secular perturbations.
The growth rates of such instabilities are related to one of two
characteristic time-scales.  The first is
the thermal time, defined as the characteristic heating time
\begin{equation}
t_{\rm th}\sim{U\Sigma\over 2Q^+}\sim{1\over\alpha\Omega}.
\end{equation}
{\it Thermal}
instabilities, in which a local patch of the flow undergoes runaway heating
or cooling, generally grow on this time scale.
The second time scale is the inflow time, i.e. the time it would
take for turbulent stresses to cause a fluid element to drift inward over a
distance comparable to its current radius,
\begin{equation}
t_{\rm inflow}\sim{R\over v}\sim{\Sigma\Omega R^2\over2H w_{r\phi}}
\sim{\Omega\over\alpha\Omega_{\rm K}^2}\left({R\over H}\right)^2.
\end{equation}
``{\it Viscous}" or {\it inflow} instabilities tend to grow on this
time scale, where I have enclosed the term ``viscous" in quotation marks here
(only) to emphasize that it is turbulent stresses, not microscopic viscosity,
that play a role here.  In geometrically thin accretion disks, where
dynamical equilibrium on the inflow time requires the angular velocity and
specific angular momentum of fluid orbits to be a function of radius and not
of time, one can write down a time-dependent diffusion equation for the
evolution of the surface mass density in the flow \citep{lyn74,lig74},
\begin{equation}
{\partial\Sigma\over\partial t}={1\over R}{\partial\over\partial R}\left[
{1\over\ell^\prime}{\partial\over\partial R}(R^2W_{R\phi})\right],
\label{eq:massdiffusion}
\end{equation}
where $\ell^\prime=(d/dR)(R^2\Omega)$ is the radial derivative of specific
angular momentum.  In this case, instabilities happen if $W_{R\phi}$ is
inversely related to the surface mass density, as this equation then
corresponds to a diffusion equation with a negative diffusion coefficient.
Perturbations in surface mass density would then tend to grow, rather than
be smoothed out, by this {\it anti}-diffusion.

\begin{figure}
\includegraphics[width=1.0\textwidth]{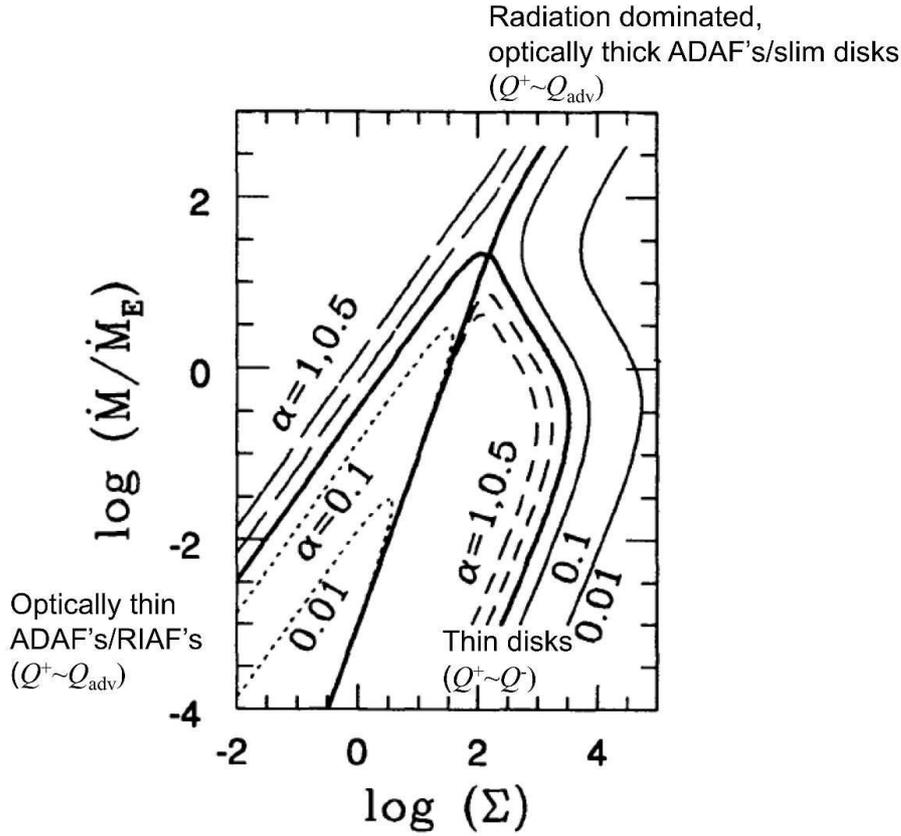}
\caption{Thermal equilibrium curves of particular accretion flow models around
a ten solar mass black hole at a particular radius $10GM/c^2$ on the
accretion rate (scaled with $\dot{M}_{\rm E}$, the Eddington luminosity
divided by $c^2$) vs. surface mass density ($\Sigma$) plane, from
\citet{che95}.  Each curve is labelled by the value of the \citet{sha73}
stress parameter $\alpha$ chosen in the model.
The locations of these curves in this plane can change considerably depending
on the physics being incorporated and how it is treated in these models - see
\citet{che95} for more details.  The topology of the curves is, however,
reasonably robust.}
\label{fig:chenetal1995}
\end{figure}

Following theoretical work on dwarf nova outbursts in cataclysmic variables
(e.g. \citealt{sma84}),
it has proved convenient to depict {\it local} thermal equilibrium
$(Q^+=Q^-+Q_{\rm adv})$ solutions
at one particular radius in a diagram of accretion rate vs. local surface mass
density.  Figure~\ref{fig:chenetal1995} \citep{che95} depicts the
topology of the space of such solutions at a radius of $10GM/c^2$ around a ten
solar mass Schwarzschild black hole.  Each curve corresponds to a different
chosen value of $\alpha$, as labelled.  The locations of these curves in
this graph can shift considerably, depending on the particular physics being
included in the models, but the topological structure is robust.  Naively,
we expect equilibrium curves with negative slopes in this diagram to be
viscously unstable, as the vertically integrated stress $W_{R\phi}$ is
proportional to the dissipation rate per unit area $Q^+$ which in turn
is proportional to the accretion rate $\dot{M}$, because, after all, it is
the inflow of matter which is the source of accretion power.  The negative
sloped curve portions in the right of this diagram near Eddington accretion
rates are radiation dominated, geometrically thin disks, and are viscously
unstable by this criterion \citep{lig74}.  They are also thermally unstable
\citep{shi75,sha76}, as can be seen because $Q^+$ exceeds (is less than)
$Q^-+Q_{\rm adv}$ above (below) these curve portions.  Hence a perturbation
upward (downward) from this curve will cause runaway heating (cooling), moving
it away from the equilibrium curve.  Similarly, the middle bold line in
the lower portion of this diagram, which corresponds to a hot, optically thin
flow where turbulent dissipation is balanced by radiative cooling is
viscously stable but thermally unstable.  This solution was first discovered
by \citet{shap76}.

This leaves three regions of the diagram which appear to correspond to thermally
and viscously stable solutions.  The lower right set of curves
correspond to the gas pressure dominated regime of the original geometrically
thin, optically thick accretion disks of \citet{sha73}.
At higher accretion rates radiation pressure starts to dominate these
geometrically thin solutions and the curves bend over to the unstable
negative slopes.  At still higher accretion rates, the inflow time becomes
shorter than the cooling time, so that heat is advected inward.
This advection is stabilizing \citep{abr88} and, because the scale height of
the disk can become quite large, these flows have been dubbed ``slim disks".
(Note that for larger radii, the optically thick, geometrically thin
disk curves would also bend toward unstable negative slopes and then back toward
positive stable slopes as one {\it lowers} the accretion rate and passes through
the transition where ionized hydrogen becomes neutral.
This unstable branch is responsible for the transient outburst behavior
observed in many black hole and neutron star X-ray binaries, as
well as dwarf novae in accreting white dwarf systems.  We will have little
to say about this instability here, as it generally occurs in the outer,
less luminous portion of the disk in accreting black hole systems, but it
is crucial for explaining the phenomenology of black hole transients.  See
\citealt{las01} for a good review of the theory.  Note, however, that these
outbursts provide some of the only constraints on the levels of turbulent
stress in accretion disks, with $\alpha\simeq0.1-0.3$ in the outburst phase.
This is significantly higher than that measured in local shearing box
simulations with no net vertical magnetic flux, as illustrated in
Figure~\ref{fig:alphavspradopgas} above.  It may be that the character of the
turbulence changes when one is so close to the regime of hydrogen ionization,
or it may be that external magnetic flux is necessary to explain the
observations.  This is a significant unsolved problem, e.g.
\citealt{kin07,kot12}.)

In addition to these two optically thick accretion disk solutions, a third
set of stable solutions exists which is optically thin and which only exists
at low accretion rates, provided $\alpha$ is not too high.  Here the cooling
time significantly exceeds the infall time, so that advection again balances
turbulent dissipation, and this apparently also stabilizes the flow
\citep{ich77, nar94, abr95, che95, nar95a, nar95b}.  Such flows are known as
Advection Dominated Accretion Flows (ADAF's, a term which can also be
applied to the optically thick, radiation dominated slim disks) or radiatively
inefficient accretion flows (RIAF's).

While the above arguments suggest that these three sets of solutions are
thermally and viscously stable, attempts to rigorously demonstrate this
involve subtle issues, particularly in the two advection dominated solutions,
and the situation is in fact not entirely clear.  Advection dominated
solutions are fairly geometrically thick ($H\sim R$), so the thermal and inflow
time scales are comparable, and there is no longer a clean separation of
thermal and viscous instabilities.  Thermal pressure is also not dynamically
negligible, so that thermal perturbations can alter the specific angular
momentum distribution and the surface mass density in the disk, even
without mass diffusion due to turbulent stresses.  Attempts to tackle this
problem have been made (e.g. \citealt{kat96,wu96,kat97,wu97,yam97}), but they
involve considerations of turbulent heat diffusion and turbulent bulk
viscosity, for which there is currently very little understanding in the
context of MRI turbulence.  One hopes that simulations will shed light on
these issues, and currently global MRI simulations in the RIAF regime appear to
be consistent with thermal and viscous stability (see Chapter 2.4).
As we discuss further below, simulations of the optically thick solutions,
which are far more challenging, currently cast doubt on alpha prescription
stability analyses, even in the supposedly unstable geometrically thin,
optically thick radiation pressure dominated solution.

It should also be noted that, in addition to the thermal and viscous
instabilities, other, shorter time scale instabilities exist that are driven
by the thermodynamics of the alpha stress prescription, particularly the
excitation of acoustic modes (e.g. \citealt{blu84,kat88,che93}),
and these might be relevant for explaining high
frequency variability in black hole sources.  Again, however, it is far
from clear that the time-dependent thermodynamics of the alpha prescription
accurately represents the time-dependent thermodynamics of MRI turbulence.
On the other hand, {\it dynamical} excitation of
nonaxisymmetric acoustic waves clearly occurs in MRI turbulence
\citep{hei09a,hei09b}.

Each of the three solutions is expected to have a distinct relationship
between overall radiative luminosity and accretion rate.  Geometrically
thin, radiatively efficient accretion disks are expected to have a
luminosity which varies linearly with accretion rate, as for a given black
hole spin, all the released
binding energy is equivalent to an approximately fixed fraction of the rest mass
energy of the accreted material.  Advective models have reduced radiative
efficiency.  For the optically thin RIAF solutions, the luminosity is
approximately proportional to the square of the accretion rate \citep{nar95b}
and therefore the radiative efficiency drops as the accretion rate is reduced.
Observational tests of this predicted relationship are discussed in Chapters
3.2 and 5.7.  In the high luminosity regime of slim
disks, the radiative output is expected to approximately saturate to the
Eddington limit as photons become trapped \citep{beg82}, and this is expected
to remain true even if outflows are driven (e.g. \citealt{pou07}), although
much of this luminosity will be emitted anistropically toward the rotation
axis (e.g. \citealt{ohs11}; Chapter 5.3).

The three basic flow paradigms have been used over the years to explain the
observed variety of black hole accretion sources.  Something
like the radiation pressure dominated advective slim disks are probably
relevant for luminous quasars and QSO's (``quasi-stellar objects"), narrow
line Seyfert 1's, ultraluminous X-ray sources, SS433, and perhaps some of the
intermediate/steep power law states of black hole X-ray binaries.  Geometrically
thin, optically thick, radiatively efficient accretion disks extending down
close to the central black holes are probably most relevant for the high/soft
state of black hole X-ray binaries and perhaps for some QSO's.  Optically thin
RIAF's are almost certainly relevant for low luminosity active galactic nuclei
(AGN), jet-dominated nonthermal AGN such as M87, the Galactic Center source
Sgr~A$^\star$, the inner regions of some broad line Seyfert 1's, and the inner
regions of the low/hard state of black hole X-ray binaries.  Many of the
other chapters in this book address how well these models work in explaining
the observed properties of these sources.

\subsection{Tying local models into global models:  the overall geometry
of the flow}

\begin{figure}
\includegraphics[width=1.0\textwidth]{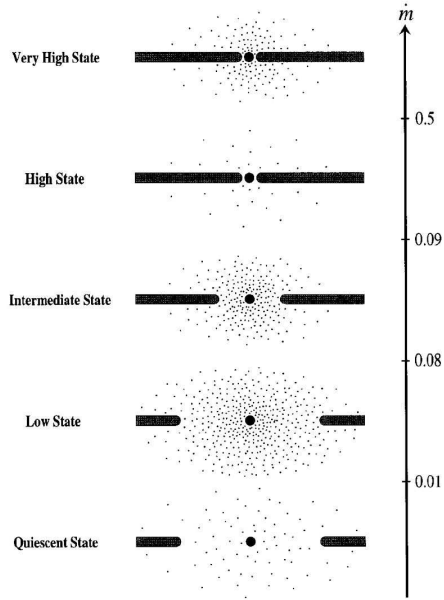}
\caption{Suggested flow geometries involving outer geometrically thin/optically
thick disks and inner optically thin RIAF's for various observed states in
black hole X-ray binaries, from \citet{esi97}.  Here $\dot{m}$ is the accretion
rate scaled by ten times the Eddington luminosity divided by $c^2$.}
\label{fig:esinetal}
\end{figure}

Figure~\ref{fig:chenetal1995} shows that, provided alpha is not too high, the
RIAF solutions at any particular radius always terminate above a critical
accretion rate where radiative cooling becomes comparable to turbulent
dissipation \citep{abr95,nar95b}.  This critical accretion rate generally
decreases with radius at large radii, so that provided the accretion rate is
not too low, the accretion flow will generally be in the radiatively efficient,
geometrically thin disk state at large radii.  However, at smaller radii,
provided the accretion rate is not too high
(${\dot M}\lesssim10\alpha^2\dot{M}_{\rm E}$, where $\dot{M}_{\rm E}$ is the
Eddington luminosity divided by $c^2$; \citealt{nar95b}), the accretion flow
can exist in one of two distinct thermal equilibria:  the
optically thin RIAF solution, and the radiatively efficient geometrically
thin disk solution.  \citet{nar95b} argue that evaporation from the surface
of the thin disk will tend to drive the accretion flow into the RIAF regime
whenever it is possible.  One would then be left with a geometry which
consists of an outer geometrically thin disk extending down toward a
transition
radius inside of which the flow adopts the RIAF solution.  As the external
accretion rate increases, the transition radius moves inward.  This then
provides an explanation for transitions between hard and soft states
in black hole X-ray binaries (\citealt{esi97}, Figure~\ref{fig:esinetal}).
As discussed in Chapter 2.5, it is now well-established that the
external accretion
rate is not the only parameter that controls state transitions:  hard to soft
transitions generally occur at higher accretion rates than soft to hard
transitions.  Nevertheless, this geometry of an outer thin disk and an inner
RIAF has become a popular model for hard states of black hole X-ray binaries,
and of certain classes of active galactic nuclei.

RIAF's are not the only way to produce hard X-rays, however.  As illustrated
in the top most panel of Figure~\ref{fig:esinetal},  it is possible that a
corona containing hot or energetic nonthermal electrons exists above and
below the geometrically thin disk.  This could be locally generated by, for
example, flares associated with buoyant magnetic field lines in a manner
analogous to the production of the solar corona \citep{gal79,haa91}.  Alpha
disk models in which some fraction of the locally generated accretion power
is dissipated in an external corona have been developed by \citet{sve94}.
Just as in the sun, the actual geometry of the corona could be quite
complicated, with multiple coronal patches.

\begin{figure}
\includegraphics[width=1.0\textwidth]{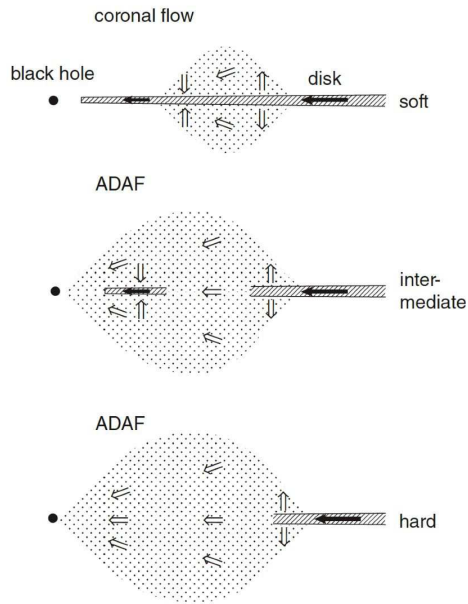}
\caption{Possible accretion flow geometries in evaporation/condensation
models, from \citet{mey07}.}
\label{fig:meyeretal}
\end{figure}

Starting with the work of \citet{mey00a,mey00b} and \citet{roz00},
many attempts have also been made to build alpha-based models of outer
thin disks, inner RIAF's, and coronae that themselves are treated as accreting
RIAF's but which can sandwich portions of the thin disk.  Each of the
different flow regions exchange energy and mass through thermal conduction,
evaporation and condensation, and irradiation.  As shown in
Figure~\ref{fig:meyeretal}, one can even form inner condensed pieces of
radiatively efficient thin disks embedded inside the RIAF/corona flow
in these models (e.g. \citealt{mey07,liu11}).  While such flow geometries
may well occur in nature, using the same alpha prescription everywhere,
especially at high latitudes off the midplane, might be problematic, as
discussed briefly in section 3.2 below.
One hopes that thermodynamically consistent global simulations of MRI
turbulence may shed light on how transitions between thin disks, coronae,
and RIAF's actually occur.

\subsection{Other variants}

In addition to the three basic local models of optically thin RIAF's,
radiatively efficient geometrically thin disks, and radiation pressure
dominated advective slim disks, numerous other models have been proposed
over the years.  Some of these are variations on the three basic models,
while others involve more significant departures from the physics included
in these models.

Because much of the dissipated accretion power is not radiated away, both
the optically thin and optically thick advection dominated solutions involve
fluid which is only weakly bound to the black hole, i.e. with internal
energy comparable in magnitude to the orbital binding energy.  Outflows are
therefore very likely to occur in these regimes \citep{nar94,bla99}, and one
way of incorporating them (with little physics beyond the invocation of
self-similarity) is to simply assume that the accretion rate varies as
a power law with radius $\dot{M}\propto R^p$ \citep{bla99}.  Assuming 
accretion velocities scale with the free-fall speed $\propto R^{-1/2}$, as
self-similarity would require, such solutions have a radial density profile
$\rho\propto R^{p-3/2}$, with $p=0$ corresponding to the standard ADAF.
Such solutions have been
dubbed Advection-Dominated Inflow-Outflow Solutions (ADIOS) by \citet{bla99}.
Outflows have been commonly observed in global simulations of these flow
regimes, (Chapters 2.4 and 5.3), though exactly how much mass is lost
compared to how much is accreted remains theoretically uncertain.  Very recent
simulations by \citet{nar12} in the low luminosity RIAF regime find
that outflows are not as powerful as previously thought.  Time-dependent
self-similar ADAF solutions by \citet{ogi99}, that differ from the stationary
self-similar solution of \citet{nar94}, appear to be consistent with
this result.  At high luminosities, outflows can
also be driven directly by radiation pressure, particularly with the high
atomic opacities expected for gas around supermassive black holes in active
galactic nuclei \citep{mur95,pro00}.

Another consequence of not dissipating accretion power in advection dominated
flows is that the entropy of the plasma increases inward, implying that the
flow will be, at least hydrodynamically, convectively unstable.  It has been
suggested that large scale convection in the flow can transport angular
momentum {\it inward}, and stationary solutions in which only small accretion
rates occur as material circulates again and again in convective eddies have
been proposed which have radial density profiles $\rho\propto R^{-1/2}$
(Convection Dominated Accretion Flows or CDAF's; \citealt{nar00,qua00}).
The physical consistency of such solutions is controversial \citep{bal02,nar02}.
The recent RIAF simulations by \citet{nar12} also do not find obvious signs of
convection.

Yet another variant on the optically thin RIAF model is the luminous hot
accretion flow (LHAF) model of \citet{yua01}.  As we discussed above, unless
alpha is large, RIAF's have a maximum accretion rate above which the solutions
with advective cooling do not exist.  This creates a problem for using them
to explain hard state sources that are observed to exist at high luminosities.
The LHAF model solves this problem by positing that advective cooling is
replaced by advective {\it heating}.  The heating here is due to compressional
work, and is non-dissipative.  LHAF's therefore essentially balance
compressional work plus turbulent dissipation with radiative cooling.  However,
such models appear to be thermally unstable given their local equilibrium curves
on the accretion rate vs. surface mass density plane, although the growth
rates might be long enough to not significantly affect the flow \citep{yua03}.

Models involving strong and/or large-scale magnetic fields in the flow
have also been proposed, for example flows in which magnetic
pressure dominates thermal pressure \citep{par03,mei05,mac06,beg07}, and
flows in which angular momentum transport is dominated by large-scale MHD
outflows (e.g.  \citealt{fer95}).  It is even possible for enough magnetic
field to be advected inward in the accretion flow that it becomes strong
enough to disrupt the flow, allowing accretion to only proceed through
magnetic Rayleigh-Taylor interchange motions (so-called Magnetically Arrested
Disks; \citealt{nar03}, see also Chapter 2.4).  Such flows are beyond the
scope of this particular review, which focuses on accretion driven by
turbulence, although they may very well be important in nature.  
It is now time for us to turn to what is known about the local
properties of MRI turbulence in accretion flows.

\section{Going beyond alpha:  MRI turbulence}

Since the discovery of its relevance to the physics of
accretion flows \citep{bal91,haw91}, tremendous theoretical effort has been
expended to try and understand the properties of MRI turbulence in various
regimes of relevance to astrophysics.  Studying MRI turbulence
has allowed us to pose questions that simply cannot be asked within the
alpha prescription, and has led to considerable new physical insight into
how the accretion process works, as well as sharpened the true physical
uncertainties.  It is, after all, only by building models based on real
physics, rather than precriptions that sweep undertainties into a single
parameter, that real scientific progress can be made.  However, it has to be
admitted that the ultimate goal of replacing alpha-based modeling of accretion
powered sources with observationally falsifiable models based on the actual
physics of the turbulence has {\it not} yet been achieved.  In this section
I will
review some of the fundamental physics issues that have been understood, or
at least revealed, by {\it local} simulations of MRI turbulence.  Global
MRI simulations of black hole accretion flows have also provided considerable
insight, and are in fact closest to realizing the goal of providing
observationally testable models of the accretion flow onto the Galactic Center
black hole source Sgr~A$^\star$.  These global simulations are reviewed in
Chapter 2.4.

\subsection{Shearing box simulations of MRI turbulence}

It is in the very nature of fully nonlinear, strong turbulence that
energy released or injected from large spatial scales passes quickly
down to microscopic dissipation scales through a turbulent cascade.  In our
case gravitational binding energy is released through the MRI which grows
by tapping directly into the free energy associated with the differential
rotation inherent in the accretion flow.  The microscopic dissipation scales
are associated with the true viscosity and resistivity of the plasma.
The actual physical dissipation scales relevant to black hole accretion
flows are extremely small compared to the energy release scales of the MRI
(presumably of order the disk thickness), but numericists have nevertheless
hoped that by putting in enough grid zones into their simulations, that some
convergence can be achieved in describing the properties of MRI turbulence.
That hope is best achieved in local shearing box simulations of the turbulence,
where all the computing power is devoted to resolving scales within the
turbulent cascade, and not on the larger scale dynamics associated with the
overall flow geometry (as important as these larger scales are to ultimately
understanding observed sources).

The geometry and properties of the shearing box are very nicely described
by \citet{haw95}.  Essentially a small, perfectly rectangular Cartesian box
is placed in the rotating shear flow, and corotates with the background flow
at the center of the box.  The curvature in the background flow streamlines
are entirely neglected, but the effects of rotation are nevertheless included
through Coriolis forces as well as centrifugal forces that are combined with
the gravitational force through an effective potential.  Boundary conditions
are such that the flow is assumed to be perfectly periodic in the azimuthal
direction, but {\it shearing} periodic in the radial direction:  one
imagines many identical shearing boxes sliding past each other according
to the background differential rotation.  If the box is placed in the midplane
of the flow, one sometimes neglects the vertical gravity and adopts periodic
boundary conditions in the vertical direction ({\it unstratified shearing
boxes}), but one can also include vertical gravity ({\it stratified shearing
boxes}) and adopt outflow boundary conditions, or, for computational
convenience, retain vertical periodic boundary conditions (a stack of
accretion disk pancakes!).

The symmetries of the standard shearing box mean that there is {\it no}
net accretion of mass through the box, and therefore in fact {\it no}
release of gravitational binding energy.  All
the energy associated with the turbulence in fact arises from the net work
done by the turbulent stresses on the shearing walls of the box.

Any initial net vertical magnetic field must be conserved in a shearing box
simulation, and this is also true of net azimuthal magnetic field if the
shearing box is unstratified (such field can leak out of the vertical
boundaries of stratified boxes if outflow boundary conditions are employed).
Shearing box simulations can therefore have the net magnetic flux through the
box as a fixed external parameter.  Unstratified simulations with no external
flux, and no explicit treatment of viscosity and resistivity which would
resolve dissipation on scales larger than the grid scale, actually produce
MRI turbulent stresses that monotonically decrease with increasing numerical
resolution \citep{pes07,fro07}!  However, this is a singular situation, as
including explicit viscosity and resistivity in the MHD equations (albeit
at far larger values
than are relevant for black hole accretion flows) or a net magnetic flux
does lead to converged levels of stress as numerical resolution is increased.
This stress increases with the amount of external magnetic flux
\citep{haw95,pes07} and also increases with the dimensionless ratio of
kinematic viscosity to Ohmic resistivity, known as the magnetic Prandtl
number \citep{les07,fro07b,sim09}.  The result that stress increases with
net magnetic flux has also been confirmed in localized regions of global
simulations \citep{sor10}.

\begin{figure}
\includegraphics[width=1.0\textwidth]{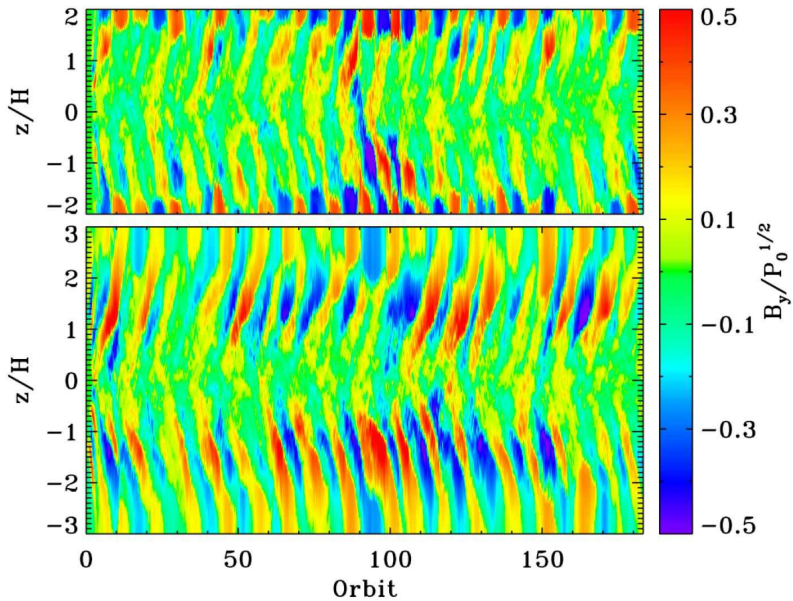}
\caption{Horizontally-averaged azimuthal component of the magnetic field as
a function of height $z$ and time in two vertically stratified shearing box
simulations whose only difference is the height of the box, from \citet{dav10}.
There exists a dynamo in such vertically stratified simulations that causes
quasiperiodic azimuthal field reversals.}
\label{fig:davisbutterfly}
\end{figure}

Adding in vertical gravity in stratified simulations enables convergence
of the turbulent stresses even without net magnetic flux and explicit
viscosity and resistivity \citep{shi10,dav10}.  A major difference between
unstratified and stratified shearing box simulations is that gravity allows
for magnetic buoyancy, and this is clearly playing a role as alternating
signs of azimuthal field continually develop in a dynamo within the
MRI turbulence and rise outward \citep{bra95}.
For weak magnetic fields, this produces
a quasiperiodic pattern of field reversals, as shown in
Figure~\ref{fig:davisbutterfly}.  (When time-reversed, the pattern here
resembles the butterfly diagram of latitudunal distributions of sunspots over
the course of the solar cycle.  By analogy, this behavior is occasionally
referred to as the MRI butterfly diagram.)
Moreover, stratified shearing box simulations with vertical outflow boundary
conditions and net vertical magnetic flux can actually locally produce
magnetocentrifugally driven outflows \citep{suz09,fro12,les12,bai12}.
Increasing the net vertical magnetic flux in such simulations can also
destroy the periodicity of the MRI butterfly dynamo, and ultimately suppress it
\citep{bai12}.

The whole premise of the alpha prescription is that the stresses are
inherently local:  (vertically-averaged) stress just depends on
(vertically-averaged) thermal pressure.  This appears to be mostly confirmed
by shearing box simulations which have wide radial extents, in that spatial
correlations between stresses at different locations decrease rapidly on
radial scales larger than the disk scale height.  However, there remain
$\sim20$~percent correlations in the Maxwell stress on larger radial scales,
indicating that the turbulence may not be entirely local \citep{sim12}.
The butterfly dynamo cycles have also been observed in global simulations
and have significant radial coherence on scales much larger than a disk scale
height \citep{one11}.

\subsection{Aspects of the vertical structure revealed by MRI simulations}

Provided any external magnetic flux is not too high, stratified shearing
boxes generally result in a structure that is dominated by thermal pressure
in the midplane and magnetic pressure and tension forces in the outer layers
\citep{sto96,mil00,hir06}.  MRI turbulence is generally confined to the
weakly magnetized regions near the midplane, while Parker instability
dynamics dominates the outer regions \citep{bla07}.  This basic structure
of weakly magnetized midplane regions and more strongly magnetized high
altitude regions is also generally observed in global simulations
(e.g. \citealt{haw02,pen10,sor10}), although
here the strongly magnetized regions at high altitude can involve significant
radial flows and circulation, which cannot happen in a shearing box.
The fact that the magnetic pressure profile is broader than
the thermal pressure profile, and that Maxwell stresses generally dominate
Reynolds stresses in the turbulence, implies that an average of the ratio
of stress to thermal pressure (alpha!) generally increases outwards:  alpha
should never be treated as a local quantity, but instead it is, at best,
a representation of the ratio of vertically-averaged stress to pressure.
Disk atmospheres are generally supported by magnetic fields, not thermal
pressure, so that if atmosphere models of thermal and reflection spectra
rely critically on vertical hydrostatic equilibrium between gravity and
thermal pressure gradients, they may not be accurate.  In addition, the fact
that MRI turbulence is generally confined to the weakly magnetized midplane
regions suggests that models of accreting coronal flows discussed in section
2.2 above are probably not well described by a simple alpha prescription.

The magnetically-dominated outer layers are very suggestive of the locally
generated magnetized corona discussed above in section 2.2. Indeed, very
tall stratified shearing box simulations by \cite{mil00} found that
approximately a quarter of the magnetic energy generated in the turbulent
midplane regions was carried out into the corona.  These simulations assumed
an isothermal equation of state, however, and vertically stratified
simulations that capture turbulent dissipation as heat and incorporate
diffusive radiation transfer have generally found that the fraction of 
accretion power that is dissipated in the magnetized corona outside the
photosphere is very small \citep{hir06,kro07,hir09b}.  On the other hand,
the magnetic buoyancy exhibited in the butterfly diagram illustrated in
Figure~\ref{fig:davisbutterfly} can be very energetically important in
transporting significant amounts of thermal energy outward in the form
of trapped photons in the radiation pressure dominated regime \citep{bla11}.
It should also be emphasized that the existing radiation MHD simulations
have not included a net vertical magnetic flux, which might in principle
enhance the coronal energetics.

\subsection{Physics issues in the RIAF regime}

Low luminosity RIAF models continue to be plagued by significant
uncertainties in the microphysics of the plasma, whether they are globally
simulated with MRI turbulence or modeled with an alpha prescription.  Because
the accreting plasma retains a significant fraction of its binding energy as
internal energy, temperatures must approach virial temperatures:
$kT\sim GMm_{\rm p}/R=(R_{\rm g}/R)m_{\rm p}c^2$, where $m_{\rm p}$ is the
proton mass.  This corresponds to $\sim10^{12}$~K at ten gravitational radii.
Optically thin cooling by electrons at such temperatures is very fast, unless
the accretion rate and density is extremely low, so
in order to not radiate away all the heat on an inflow time, the electron
temperature $T_{\rm e}$ must be significantly less than the ion temperature
$T_{\rm i}$.  This in
turn implies that the electrons should not receive the vast majority of
the turbulent dissipation of accretion power (otherwise they would
radiate it away), and the ions must be at least partially thermally decoupled
from the electrons on the inflow time.  Coulomb collisions alone will be
insufficient to thermally couple ions and electrons provided the accretion
rate is not too high \citep{ree82,nar95b}.  Plasma instabilities
may in principle exist that couple the species more rapidly, but so far
this has not been demonstrated \citep{beg88,par10}.

Even the MRI behaves
differently in the collisionless regime \citep{qua02,sha03,bal04,isl05},
especially in giving
rise to anisotropic pressure tensors that themselves can give rise to
significant angular momentum transport \citep{sha06}, and this is not captured
in simulations that assume MHD.  How the turbulent dissipation is ultimately
channeled into
heating of the ions and electrons (or energization - the ion and electron
distribution functions need not be thermal) is another significant
uncertainty.  Local simulations
find that direct heating by the anisotropic pressure tensor
can account for 50 percent of the heating by the turbulence, and that the
ratio of electron to ion heating is $\sim0.3(T_{\rm e}/T_{\rm i})^{1/2}$
\citep{sha07}.  It is also just
now becoming possible to do fully kinetic simulations of collisionless
MRI turbulence, at least locally \citep{riq12}, so that there is hope for
further resolving some of these issues in the not so distant future.

All these effects remain to be included in global simulations of the RIAF
regime, discussed in Chapter 2.4.  Currently these generally assume regular MHD,
and adopt prescriptions for treating the electron and ion distribution
functions, such as assuming thermal distributions with a constant ion to
electron temperature ratio.

\subsection{Attempts to simulate the radiation pressure thermal instability}

For some years now, it has been possible to do vertically stratified shearing
box simulations that explore the thermodynamics of MRI turbulence.  These
simulations capture grid scale losses of magnetic and kinetic energy, and
incorporate radiation transport and cooling through flux limited diffusion.
It has also been possible to do global simulations using flux limited
diffusion under axisymmetry - see Chapter 5.3.
Recently, even more accurate radiation transport algorithms have been
successfully developed (e.g. \citealt{dav12,jia12,sad12,tak12}).

Vertically stratified shearing box simulations with optically thick cooling
and which incorporate the dynamics of radiation pressure enable the exploration
of the thermal instability predicted by alpha disk modeling on the negative
slope branch of the thermal equilibrium curves on the right hand side of
Figure~\ref{fig:chenetal1995}.  The origin of this thermal instability is
very easy to understand.  Assuming that radiation diffusion dominates the
vertical heat transport (which is generally found in simulations), the local
cooling rate per unit area is of order the radiation energy density at the
midplane $aT^4$ times the speed of light over the vertical optical depth.
Because the opacity is dominated by electron scattering in these high
temperature regimes, this then implies that $Q^-\propto T^4/\Sigma$. On the
other hand, the heating rate per unit area is the vertically integrated stress
times the rate of strain, $Q^+\sim Hw_{R\phi}R|d\Omega/dR|$.  Because the disk
is supported vertically by radiation pressure, and the vertical gravity
increases linearly with height above the midplane, the disk half thickness 
is simply proportional to the surface radiation flux, i.e.
$H\propto Q^-\propto T^4/\Sigma$.  Hence a standard alpha prescription, in
which $w_{R\phi}=\alpha P=\alpha aT^4/3$, will mean that the heating rate
$Q^+\propto T^8/\Sigma$.  Because the inflow time is much longer than the
thermal time for geometrically thin disks, the surface mass density $\Sigma$
cannot vary significantly on the thermal time scale, and the heating rate
therefore depends much more sensitively on temperature than the cooling rate.
Hence a perturbative increase (decrease) in temperature would lead to runaway
heating (cooling).

As shown in Figure~\ref{fig:alphavspradopgas}, simulations are consistent
with the standard alpha prescription that the time and space averaged stress
scales with total thermal pressure, which is mostly radiation pressure in
this regime.  Nevertheless, at least some simulations have been
able to establish thermal equilibria between heating and cooling in the
radiation pressure dominated regime that last for many thermal times
\citep{tur04,hir09b}.  On the other hand, recent simulations by
\citet{jia13b} find that such equilibria, even if established, always
eventually suffer runaway heating or cooling.  It is not clear what is
producing the differences between the different simulations, which are
run using different codes.

\begin{figure}
\includegraphics[width=1.0\textwidth]{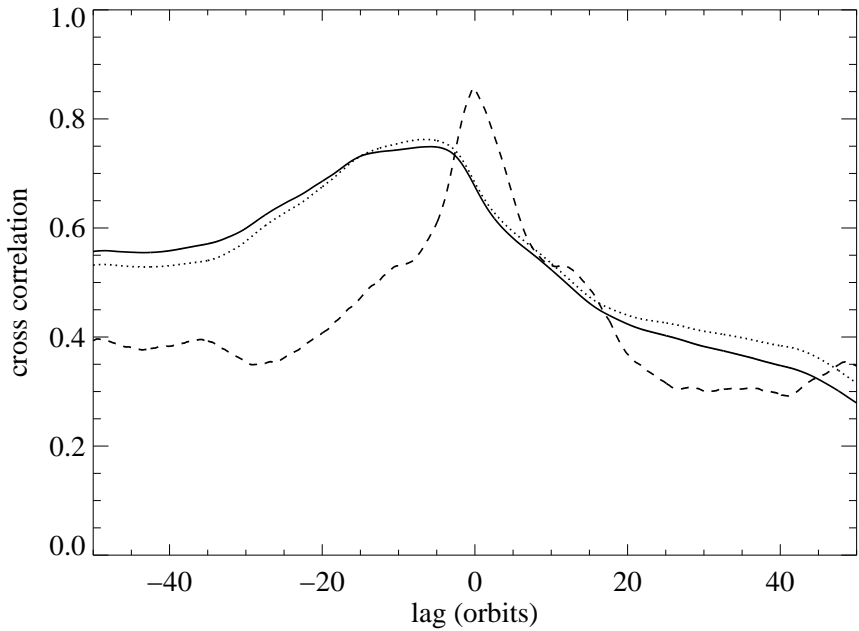}
\caption{Cross correlation coefficient for various forms of volume
integrated energy as a function of time difference with respect to variations
in volume integrated magnetic energy, for a radiation dominated stratified
shearing box simulation.  Negative values on the horizontal axis
mean that the energy lags behind magnetic energy.  The dashed line shows the
turbulent kinetic energy, which is highly correlated with magnetic energy
at zero lag.  This is because both energies are aspects of the same MRI
turbulence!  The solid and dotted curves show radiation and gas internal
energies, respectively.
Both of these are very similar as they are both thermal energies, and both
are again correlated with magnetic energy, but with a significant time lag
of order 5-15 orbital periods.  This is comparable to the thermal time in
this simulation.  (From \citealt{hir09b}.)}
\label{fig:hirosecrosscorr}
\end{figure}

Note that, in contrast to the hydrogen ionization driven thermal/viscous
instability that is responsible for dwarf nova and outbursts in X-ray binaries,
there is no similar observational evidence for the putative radiation
pressure driven thermal instability predicted by alpha disk theory.
It could be that the intrinsically stochastic nature of the turbulent
dissipation, which is not captured in the alpha prescription, acts as
a stabilizing influence \citep{jan12}.
But in addition, it is clear that the alpha prescription breaks
down on short time scales.  As shown in Figure~\ref{fig:hirosecrosscorr},
fluctuations in thermal energy {\it lag} fluctuations in turbulent energy
by approximately a thermal time.  This is easy to understand physically:
it is the dissipation of turbulence that heats the plasma and produces
thermal pressure, and thermal energy will therefore only respond to
fluctuations in turbulent energy on time scales of order the heating time
$\sim(\alpha\Omega)^{-1}$.  There appears to be no direct feedback from
pressure to stress on the thermal time scale, and the alpha prescription
is therefore only established on longer time scales, although exactly
{\it why} the prescription is established remains mysterious.  More thought
needs to be applied to understanding the relationship between stress
and pressure, and to understanding what is going on to explain the different
results seen in different simulation codes.  Recently,
\citet{lin12} have taken a useful first step
in developing analytic, time-dependent ordinary differential equations that
successfully reproduce a number of the observed features in these simulations.

Note that while the thermal instability has been studied using local
simulations of MRI turbulence, the inflow or viscous instability
has not as the boxes that have been used are radially too narrow to
allow for significant variations in surface mass density.  Very wide
radial boxes, or global simulations, will be necessary to explore this
physics.

\subsection{Other issues in the radiation dominated regime}

Radiation dominated plasmas have a number of other interesting properties
that are likely relevant to black hole accretion flows in the high
luminosity regime.  One in particular is that the sound speed in the
fluid $\simeq[4aT^4/(9\rho)]^{1/2}$ is determined by radiation pressure,
not gas pressure, and when the former exceeds the latter, it is possible
to be in a regime where turbulent motions are subsonic and yet supersonic
with respect to the sound speed from gas pressure alone.
But photons generally diffuse through the
plasma, and if they do so rapidly, then even fluid motions that are
subsonic with respect to the radiation pressure sound speed, but supersonic
with respect to the gas sound speed, can be highly compressible, because
photon diffusion reduces the photon pressure response.  Large density
fluctuations can therefore be produced, and this
has been observed in unstratified shearing box simulations of
radiation dominated MRI turbulence \citep{tur03,jia13a}.  

Radiation damping of temperature fluctuations in radiation pressure
dominated MRI turbulence can be a significant source of dissipation which,
like the pressure anisotropies in collisionless MRI turbulence discussed
above in section 3.3, can be resolved in numerical simulations.  Such
fluctuations can be compressible in nature as we just mentioned
\citep{ago98}, or due to nonlinear isobaric fluctuations associated
with regions of high magnetic pressure \citep{bla11}.  Some tens of percent
of the total dissipation has been observed to occur through radiation damping
in shearing box simulations \citep{tur04,bla11}.  This radiation damping can
also increase the bulk viscosity and therefore the magnetic Prandtl number,
and can increase the Maxwell stress in the turbulence \citep{jia13a}.

The fact that MRI turbulence can be supersonic with respect to the sound speed
in the gas alone in the radiation dominated regime implies that, in principle,
turbulent speeds can exceed mean thermal speeds not only of the ions in
the plasma, but also the electrons.  If turbulent motions are limited by
the radiation sound speed then this may start to happen at
radiation to gas pressure ratios in excess of the proton to electron mass
ratio, and may happen at even lower ratios in the photosphere regions
which tend to be dominated by magnetic pressure, not thermal pressure.
Differences in bulk turbulent velocities on the scale of a photon mean free
path that exceed in magnitude characteristic electron thermal speeds will
mean that bulk Comptonization by the turbulence itself will dominate
thermal Comptonization, and this may provide an alternative means of
producing a Comptonized high energy spectrum in radiation dominated
luminous states of black hole sources \citep{soc04,soc10}.

\begin{figure}
\includegraphics[width=1.0\textwidth]{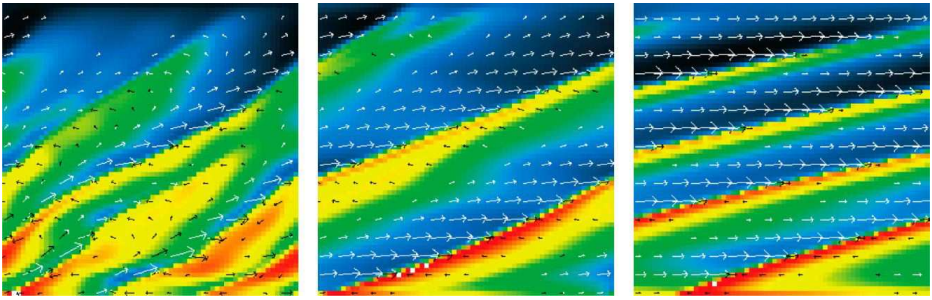}
\caption{2D simulations of shock trains produced by the photon bubble
instability in radiation pressure supported media with initially uniform
magnetic fields of increasing strength from left to right.  Arrows show
fluid velocity and colors show density on a logarithmic scale, with warm
colors being high and cold colors being low.  The weight of the dense
shocked fluid causes the weaker magnetic fields in the left hand figure
to buckle.  (From \citealt{tur05}.)}
\label{fig:turnershocks}
\end{figure}

Advection of heat is a key ingredient to the radiation pressure
dominated slim disk solutions discussed in section 2.1 above. However,
instabilities might produce inhomogeneities in the flow that allow photons
to escape more readily through underdense channels, rather than be
advected inward.  The most well-explored of such
instabilities are magnetically-mediated ``photon bubble" instabilities
\citep{aro92,gam98}.  On short length scales where photons diffuse
rapidly, such instabilities amount to radiatively amplified magnetosonic
modes \citep{bla03} that develop into highly inhomogeneous
trains of shocks (\citealt{beg01,tur05}, see Figure~\ref{fig:turnershocks}).
In principle such inhomogeneities could allow locally super-Eddington fluxes
to escape from the disk atmosphere without driving an outflow \citep{beg02}.
Simulating photon bubbles in the presence of MRI turbulence has proved
computationally challenging in the radiation pressure dominated regime
due partly to the small length scales (of order the {\it gas} pressure scale
height) that must be resolved, and partly by the fact that Parker
instabilities in the magnetically dominated surface layers also produce
significant inhomogeneity.  Models of slim disks with porous
outer layers {\it and} winds have recently been developed by \citet{dot11}.
It has also been suggested that accretion flows
in the radiation dominated regime might be highly inhomogeneous structures
that are not well-described by any of the standard accretion flow models
discussed in section 2.1 \citep{dex11}.

\section{Summary}

The primary reason why the alpha prescription continues to be a mainstay
of black hole accretion theory is that it enables models to be built that
couple the dynamics of the flow (outward angular momentum transport through
the plasma by a stress described phenomenologically by alpha) to the
thermodynamics (dissipation of accretion power described phenomenologically
by the same alpha stress prescription times the shear rate).  Once this is all
combined with radiative cooling processes, one has models that can be used to
generate spectra, time variability, etc. that can be compared to
observations.  While decades of theoretical work within this research
paradigm have yielded valuable insights (e.g. scalings of luminosity
and temperature with accretion rate, the importance of advection of heat),
one can only carry this program so far without addressing the fundamental
physics that the alpha prescription hides.

(Fewer) decades of research have now been spent on understanding the
properties of MRI turbulence, but until recently, this has focused
mostly on the dynamics of the turbulence, not the thermodynamics, and it
is the latter which is ultimately required to connect to observations,
which are, after all, detecting the photons emitted by the source.  We still
have a lot of unanswered fundamental questions.  For example, what is
the true nature of the thin disk/RIAF transition radius, upon which hangs so
much phenomenology of black hole X-ray binaries (hard/soft state transitions,
band-limited noise, low frequency quasi-periodic oscillations, ...)?  The
thermal/viscous instabilities have guided theoretical effort to exclude
uphysical equilibria, but are these instabilities always real, and how do
they manifest?  (Only those driven by ionization/recombination are
observationally known to exist, e.g., in dwarf novae.)
What determines how accretion power is partitioned into
various forms?  These are all questions of thermodynamics, not just dynamics.

As discussed in Chapter 2.4, global MRI simulations of low luminosity
RIAF's have been most successful in connecting to observations, largely
because, at least in some regimes, the radiative cooling does not dramatically
affect the dynamics of the flow, and therefore can be calculated after the
fact by post-processing the simulation.  Moreover, because the flow is
optically thin at most photon frequencies, the simulation
hardware has actually developed to the point where radiative cooling can be
fully incorporated in the actual dynamical simulation itself \citep{dib12}.
However, as
discussed in section 3.3 above, even here there remain issues in the
microphysics that still need to be understood theoretically.

Optically thick radiation transport is computationally more expensive, and
also more important for the dynamics in the high luminosity
black hole accretion regimes where radiation pressure plays a critical role.
As discussed in section 3.4 above, local shearing box simulations have shed
some light on how MRI turbulence works in this regime, and in what ways
the alpha prescription does, and does not, describe the physics.  Unfortunately,
this still leaves major unanswered questions as to the global structure
of the flow.  Two dimensional (axisymmetric) global simulations have also
been done which have confirmed the existence of discrete flow states
(\citealt{ohs09,ohs11}, see also Chapters 2.4 and 5.3), but these cannot be run
over long time scales as MRI turbulence cannot be sustained in axisymmetry.
With the ongoing increase of computer power, combined with the development of
new radiation transport algorithms, it should
be possible to do global 3D simulations of accretion flows in optically
thick regimes too.  This will better enable us to understand the origins of
state transitions in black hole X-ray binaries, transitions between radiatively
efficient thin disks and RIAF's, and other fundamental problems in
black hole accretion flows.  Achieving this goal should complete the shift
to a more physics-based research paradigm.

\begin{acknowledgements}
The author is grateful to ISSI and to the organizers of this workshop for
enabling such a productive set of scientific interactions, and thanks a
number of the participants, particularly Chris Fragile, Tom Maccarone, Shin
Mineshige, Ken Ohsuga, Juri Poutanen, and Chris Reynolds, for enlightening
him on various issues of relevance to this paper.  The author is also
grateful to the referees for suggestions that significantly improved the
manuscript.  The author's research
is supported by the US National Science Foundation under grant AST-0707624.
\end{acknowledgements}

\bibliographystyle{aps-nameyear}      
\bibliography{example}   

\begin{thebibliography}{}
%
%
\bibitem[\protect\citeauthoryear{Abramowicz et al.}{1988}]{abr88}
M. A. Abramowicz, B. Czerny, J. P. Lasota, E. Szuszkiewicz, ApJ,
\textbf{332}, 646-658 (1988)
%
\bibitem[\protect\citeauthoryear{Abramowicz et al.}{1995}]{abr95}
M. A. Abramowicz, X. Chen, S. Kato, J. P. Lasota, O. Regev, ApJ,
\textbf{438}, L37-L39 (1995)
%
\bibitem[\protect\citeauthoryear{Agol \& Krolik}{1998}]{ago98}
E. Agol, J. Krolik, ApJ, \textbf{507}, 304-315 (1998)
%
\bibitem[\protect\citeauthoryear{Arons}{1992}]{aro92}
J. Arons, ApJ, \textbf{388}, 561-578 (1992)
%
\bibitem[\protect\citeauthoryear{Bai \& Stone}{2012}]{bai12}
X.-N. Bai, J. M. Stone, ApJ, submitted (2012), arXiv:1210.6661
%
\bibitem[\protect\citeauthoryear{Balbus}{2004}]{bal04}
S. A. Balbus, ApJ, \textbf{616}, 857-864 (2004)
%
\bibitem[\protect\citeauthoryear{Balbus \& Hawley}{1991}]{bal91}
S. A. Balbus, J. F. Hawley, ApJ, \textbf{376}, 214-222 (1991)
%
\bibitem[\protect\citeauthoryear{Balbus \& Hawley}{1992}]{bal92}
S. A. Balbus, J. F. Hawley, ApJ, \textbf{400}, 610-621 (1992)
%
\bibitem[\protect\citeauthoryear{Balbus \& Hawley}{1998}]{bal98}
S. A. Balbus, J. F. Hawley, Rev. Mod. Phys., \textbf{70}, 1-53 (1998)
%
\bibitem[\protect\citeauthoryear{Balbus \& Hawley}{2002}]{bal02}
S. A. Balbus, J. F. Hawley, ApJ, \textbf{573}, 749-753 (2002)
%
\bibitem[\protect\citeauthoryear{Balbus \& Papaloizou}{1999}]{bal99}
S. A. Balbus, J. C. B. Papaloizou, ApJ, \textbf{521}, 650-658 (1999)
%
\bibitem[\protect\citeauthoryear{Begelman}{2001}]{beg01}
M. C. Begelman, ApJ, \textbf{551}, 897-906 (2001)
%
\bibitem[\protect\citeauthoryear{Begelman}{2002}]{beg02}
M. C. Begelman, ApJ, \textbf{568}, L97-L100 (2002)
%
\bibitem[\protect\citeauthoryear{Begelman \& Chiueh}{1988}]{beg88}
M. C. Begelman, T. Chiueh, ApJ, \textbf{332}, 872-890 (1988)
%
\bibitem[\protect\citeauthoryear{Begelman \& Meier}{1982}]{beg82}
M. C. Begelman, D. L. Meier, ApJ, \textbf{253}, 873-896 (1982)
%
\bibitem[\protect\citeauthoryear{Begelman \& Pringle}{2007}]{beg07}
M. C. Begelman, J. E. Pringle, MNRAS, \textbf{375}, 1070-1076 (2007)
%
\bibitem[\protect\citeauthoryear{Blaes \& Socrates}{2003}]{bla03}
O. Blaes, A. Socrates, ApJ, \textbf{596}, 509-537 (2003)
%
\bibitem[\protect\citeauthoryear{Blaes, Hirose \& Krolik}{2007}]{bla07}
O. Blaes, S. Hirose, J. H. Krolik, ApJ, \textbf{664}, 1057-1071 (2007)
%
\bibitem[\protect\citeauthoryear{Blaes et al.}{2011}]{bla11}
O. Blaes, J. H. Krolik, S. Hirose, N. Shabaltas, ApJ, \textbf{733}, 110, 24 pp.
(2011)
%
\bibitem[\protect\citeauthoryear{Blandford \& Begelman}{1999}]{bla99}
R. D. Blandford, M. C. Begelman, MNRAS, \textbf{303}, L1-L5 (1999)
%
\bibitem[\protect\citeauthoryear{Blandford \& Payne}{1982}]{bla82}
R. D. Blandford, D. G. Payne, MNRAS, \textbf{199}, 883-903 (1982)
%
\bibitem[\protect\citeauthoryear{Blumenthal, Yang \& Lin}{1984}]{blu84}
G. R. Blumenthal, L. T. Yang, D. N. C. Lin, ApJ, \textbf{287}, 774-784 (1984)
%
\bibitem[\protect\citeauthoryear{Brandenburg et al.}{1995}]{bra95}
A. Brandenburg, \AA. Nordlund, R. F. Stein, U. Torkelsson, ApJ,
\textbf{446}, 741-754 (1995)
%
\bibitem[\protect\citeauthoryear{Chen \& Taam}{1993}]{che93}
X. Chen, R. E. Taam, ApJ, \textbf{412}, 254-266 (1993)
%
\bibitem[\protect\citeauthoryear{Chen et al.}{1995}]{che95}
X. Chen, M. A. Abramowicz, J.-P. Lasota, R. Narayan, I. Yi,
ApJ, \textbf{443}, L61-L64 (1995)
%
\bibitem[\protect\citeauthoryear{Davis, Stone \& Jiang}{2012}]{dav12}
S. W. Davis, J. M. Stone, Y.-F. Jiang, ApJS, \textbf{199}, 9, 19pp. (2012)
%
\bibitem[\protect\citeauthoryear{Davis, Stone \& Pessah}{2010}]{dav10}
S. W. Davis, J. M. Stone, M. E. Pessah, ApJ, \textbf{713}, 52-65 (2010)
%
\bibitem[\protect\citeauthoryear{Dexter \& Agol}{2011}]{dex11}
J. Dexter, E. Agol, ApJ, \textbf{727}, L24, 5 pp. (2011)
%
\bibitem[\protect\citeauthoryear{Dibi et al.}{2012}]{dib12}
S. Dibi, S. Drappeau, P. C. Fragile, S. Markoff, J. Dexter, MNRAS,
\textbf{426}, 1928-1939 (2012)
%
\bibitem[\protect\citeauthoryear{Dotan \& Shaviv}{2011}]{dot11}
C. Dotan, N. J. Shaviv, MNRAS, \textbf{413}, 1623-1632 (2011)
%
\bibitem[\protect\citeauthoryear{Esin, McClintock \& Narayan}{1997}]{esi97}
A. A. Esin, J. E. McClintock, R. Narayan, ApJ, \textbf{489}, 865-889 (1997)
%
\bibitem[\protect\citeauthoryear{Ferreira \& Pelletier}{1995}]{fer95}
J. Ferreira, G. Pelletier, A\&A, \textbf{295}, 807-832 (1995)
%
\bibitem[\protect\citeauthoryear{Fragile \& Blaes}{2008}]{fra08}
P. C. Fragile, O. M. Blaes, ApJ, \textbf{687}, 757-766 (2008)
%
\bibitem[\protect\citeauthoryear{Fromang et al.}{2012}]{fro12}
S. Fromang, H. N. Latter, G. Lesur, G. I. Ogilvie, A\&A, submitted (2012),
arXiv:1210.6664
%
\bibitem[\protect\citeauthoryear{Fromang \& Papaloizou}{2007}]{fro07}
S. Fromang, J. Papaloizou, A\&A, \textbf{476}, 1113-1122 (2007)
%
\bibitem[\protect\citeauthoryear{Fromang et al.}{2007}]{fro07b}
S. Fromang, J. Papaloizou, G. Lesur, T. Heinemann, A\&A, \textbf{476},
1123-1132 (2007)
%
\bibitem[\protect\citeauthoryear{Galeev, Rosner \& Vaiana}{1979}]{gal79}
A. A. Galeev, R. Rosner, G. S. Vaiana, ApJ, \textbf{229}, 318-326 (1979)
%
\bibitem[\protect\citeauthoryear{Gammie}{1998}]{gam98}
C. F. Gammie, MNRAS, \textbf{297}, 929-935 (1998)
%
\bibitem[\protect\citeauthoryear{Gammie}{1999}]{gam99}
C. F. Gammie, ApJ, \textbf{522}, L57-L60 (1999)
%
\bibitem[\protect\citeauthoryear{Goodman}{2003}]{goo03}
J. Goodman, MNRAS, \textbf{339}, 937-948 (2003)
%
\bibitem[\protect\citeauthoryear{Haardt \& Maraschi}{1991}]{haa91}
F. Haardt, L. Maraschi, ApJ, \textbf{380}, L51-L54 (1991)
%
\bibitem[\protect\citeauthoryear{Hawley \& Balbus}{1991}]{haw91}
J. F. Hawley, S. A. Balbus, ApJ, \textbf{376}, 223-233 (1991)
%
\bibitem[\protect\citeauthoryear{Hawley \& Balbus}{2002}]{haw02}
J. F. Hawley, S. A. Balbus, ApJ, \textbf{573}, 738-748 (2002)
%
\bibitem[\protect\citeauthoryear{Hawley, Gammie \& Balbus}{1995}]{haw95}
J. F. Hawley, C. F. Gammie, S. A. Balbus, ApJ, \textbf{440}, 742-763 (1995)
%
\bibitem[\protect\citeauthoryear{Heinemann \& Papaloizou}{2009a}]{hei09a}
T. Heinemann, J. C. B. Papaloizou, MNRAS, \textbf{397}, 52-63 (2009)
%
\bibitem[\protect\citeauthoryear{Heinemann \& Papaloizou}{2009b}]{hei09b}
T. Heinemann, J. C. B. Papaloizou, MNRAS, \textbf{397}, 64-74 (2009)
%
\bibitem[\protect\citeauthoryear{Hirose, Blaes \& Krolik}{2009}]{hir09a}
S. Hirose, O. Blaes, J. H. Krolik, ApJ, \textbf{704}, 781-788 (2009)
%
\bibitem[\protect\citeauthoryear{Hirose, Krolik \& Blaes}{2009}]{hir09b}
S. Hirose, J. H. Krolik, O. Blaes, ApJ, \textbf{691}, 16-31 (2009)
%
\bibitem[\protect\citeauthoryear{Hirose, Krolik \& Stone}{2006}]{hir06}
S. Hirose, J. H. Krolik, J. M. Stone, ApJ, \textbf{640}, 901-917 (2006)
%
\bibitem[\protect\citeauthoryear{Ichimaru}{1977}]{ich77}
S. Ichimaru, ApJ, \textbf{214}, 840-855 (1977)
%
\bibitem[\protect\citeauthoryear{Islam \& Balbus}{2005}]{isl05}
T. Islam, S. Balbus, ApJ, \textbf{633}, 328-333 (2005)
%
\bibitem[\protect\citeauthoryear{Janiuk \& Misra}{2012}]{jan12}
A. Janiuk, R. Misra, A\&A, \textbf{540}, A114, 6pp. (2012)
%
\bibitem[\protect\citeauthoryear{Jiang, Stone \& Davis}{2012}]{jia12}
Y.-F. Jiang, J. M. Stone, S. W. Davis, ApJS, \textbf{199}, 14, 29 pp. (2012)
%
\bibitem[\protect\citeauthoryear{Jiang, Stone \& Davis}{2013a}]{jia13a}
Y.-F. Jiang, J. M. Stone, S. W. Davis, ApJ, \textbf{767}, 148, 14 pp. (2013a)
%
\bibitem[\protect\citeauthoryear{Jiang, Stone \& Davis}{2013b}]{jia13b}
Y.-F. Jiang, J. M. Stone, S. W. Davis, ApJ, submitted (2013b)
%
\bibitem[\protect\citeauthoryear{Kato, Abramowicz \& Chen}{1996}]{kat96}
S. Kato, M. A. Abramowicz, X. Chen, PASJ, \textbf{48}, 67-75 (1996)
%
\bibitem[\protect\citeauthoryear{Kato, Honma \& Matsumoto}{1988}]{kat88}
S. Kato, F. Honma, R. Matsumoto, MNRAS, \textbf{231}, 37-48 (1988)
%
\bibitem[\protect\citeauthoryear{Kato et al.}{1997}]{kat97}
S. Kato, T. Yamasaki, M. A. Abramowicz, X. Chen, PASJ, \textbf{49}, 221-225
(1997)
%
\bibitem[\protect\citeauthoryear{King, Pringle \& Livio}{2007}]{kin07}
A. R. King, J. E. Pringle, M. Livio, MNRAS, \textbf{376}, 1740-1746
(2007)
%
\bibitem[\protect\citeauthoryear{Kotko \& Lasota}{2012}]{kot12}
I. Kotko, J.-P. Lasota, A\&A, \textbf{545}, A115, 9pp. (2012)
(2007)
%
\bibitem[\protect\citeauthoryear{Krolik}{1999}]{kro99}
J. H. Krolik, ApJ, \textbf{515}, L73-L76 (1999)
%
\bibitem[\protect\citeauthoryear{Krolik, Hirose \& Blaes}{2007}]{kro07}
J. H. Krolik, S. Hirose, O. Blaes, ApJ, \textbf{664}, 1045-1056 (2007)
%
\bibitem[\protect\citeauthoryear{Lesur, Ferreira \& Ogilvie}{2012}]{les12}
G. Lesur, J. Ferreira, G. I. Ogilvie, A\&A, submitted (2012), arXiv:1210.6660
%
\bibitem[\protect\citeauthoryear{Lesur \& Longaretti}{2007}]{les07}
G. Lesur, P.-Y. Longaretti, MNRAS, \textbf{378}, 1471-1480 (2007)
%
\bibitem[\protect\citeauthoryear{Lasota}{2001}]{las01}
J.-P. Lasota, New Astron. Rev., \textbf{45}, 449-508 (2001)
%
\bibitem[\protect\citeauthoryear{Lightman \& Eardley}{1974}]{lig74}
A. P. Lightman, D. M. Eardley, ApJ, \textbf{187}, L1-L3 (1974)
%
\bibitem[\protect\citeauthoryear{Lin et al.}{2012}]{lin12}
D.-B. Lin, W.-M. Gu, T. Liu, M.-Y. Sun, J.-F. Lu, ApJ, \textbf{761}, 29, 5 pp.
(2012)
%
\bibitem[\protect\citeauthoryear{Liu, Done \& Taam}{2011}]{liu11}
B. F. Liu, C. Done, R. E. Taam, ApJ, \textbf{726}, 10, 5 pp.  (2011)
%
\bibitem[\protect\citeauthoryear{Lynden-Bell \& Pringle}{1974}]{lyn74}
D. Lynden-Bell, J. E. Pringle, MNRAS, \textbf{168}, 603-637 (1974)
%
\bibitem[\protect\citeauthoryear{Machida, Nakamura, \& Matsumoto}{2006}]{mac06}
M. Machida, K. E. Nakamura, R. Matsumoto, PASJ, \textbf{58}, 193-202 (2006)
%
\bibitem[\protect\citeauthoryear{Meier}{2005}]{mei05}
D. L. Meier, Astrophys. Space Sci., \textbf{300}, 55-65 (2005)
%
\bibitem[\protect\citeauthoryear{Meyer, Liu, \& Meyer-Hofmeister}{2000a}]
{mey00a} F. Meyer, B. F. Liu, E. Meyer-Hofmeister, A\&A, \textbf{354}, L67-L70
(2000)
%
\bibitem[\protect\citeauthoryear{Meyer, Liu, \& Meyer-Hofmeister}{2000b}]
{mey00b} F. Meyer, B. F. Liu, E. Meyer-Hofmeister, A\&A, \textbf{361}, 175-188
(2000)
%
\bibitem[\protect\citeauthoryear{Meyer, Liu, \& Meyer-Hofmeister}{2007}]
{mey07} F. Meyer, B. F. Liu, E. Meyer-Hofmeister, A\&A, \textbf{463}, 1-9
(2007)
%
\bibitem[\protect\citeauthoryear{Miller \& Stone}{2000}]{mil00}
K. A. Miller, J. M. Stone, ApJ, \textbf{534}, 398-419 (2000)
%
\bibitem[\protect\citeauthoryear{Murray et al.}{1995}]{mur95}
N. Murray, J. Chiang, S. A. Grossman, G. M. Voit, ApJ, \textbf{451}, 498-509
(1995)
%
\bibitem[\protect\citeauthoryear{Narayan, Igumenshchev \& Abramowicz}{2000}]
{nar00} R. Narayan, I. V. Igumenshchev, M. A. Abramowicz, ApJ, \textbf{539},
798-808 (2000)
%
\bibitem[\protect\citeauthoryear{Narayan, Igumenshchev \& Abramowicz}{2003}]
{nar03} R. Narayan, I. V. Igumenshchev, M. A. Abramowicz, PASJ, \textbf{55},
L69-L72 (2003)
%
\bibitem[\protect\citeauthoryear{Narayan et al.}{2002}]{nar02}
R. Narayan, E. Quataert, I. V. Igumenshchev, M. A. Abramowicz, ApJ,
\textbf{577}, 295-301 (2002)
%
\bibitem[\protect\citeauthoryear{Narayan et al.}{2012}]{nar12}
R. Narayan, A. Sadowski, R. F. Penna, A. K. Kulkarni, MNRAS, \textbf{426},
3241-3259 (2012)
%
\bibitem[\protect\citeauthoryear{Narayan \& Yi}{1994}]{nar94}
R. Narayan, I. Yi, ApJ, \textbf{428}, L13-L16 (1994)
%
\bibitem[\protect\citeauthoryear{Narayan \& Yi}{1995a}]{nar95a}
R. Narayan, I. Yi, ApJ, \textbf{444}, 231-243 (1995a)
%
\bibitem[\protect\citeauthoryear{Narayan \& Yi}{1995b}]{nar95b}
R. Narayan, I. Yi, ApJ, \textbf{452}, 710-735 (1995b)
%
\bibitem[\protect\citeauthoryear{Noble, Krolik \& Hawley }{2010}]{nob10}
S. C. Noble, J. H. Krolik, J. F. Hawley, ApJ, \textbf{711}, 959-973 (2010)
%
\bibitem[\protect\citeauthoryear{Ogilvie}{1999}]{ogi99}
G. I. Ogilvie, MNRAS, \textbf{306}, L9-L13 (1999)
%
\bibitem[\protect\citeauthoryear{Ohsuga \& Mineshige}{2011}]{ohs11}
K. Ohsuga, S. Mineshige, ApJ, \textbf{736}, 2, 18 pp. (2011)
%
\bibitem[\protect\citeauthoryear{Ohsuga et al.}{2009}]{ohs09}
K. Ohsuga, S. Mineshige, M. Mori, Y. Kato, PASJ, \textbf{61}, L7-L11 (2009)
%
\bibitem[\protect\citeauthoryear{O'Neill et al.}{2011}]{one11}
S. M. O'Neill, C. S. Reynolds, M. C. Miller, K. A. Sorathia,
ApJ, \textbf{736}, 107, 7 pp. (2011)
%
\bibitem[\protect\citeauthoryear{Pariev, Blackman \& Boldyrev}{2003}]{par03}
V. I. Pariev, E. G. Blackman, S. A. Boldyrev, A\&A, \textbf{407},
403-421 (2003)
%
\bibitem[\protect\citeauthoryear{Park et al.}{2010}]{par10}
J. Park, C. Ren, E. G. Blackman, X. Kong, Phys. Plasmas, \textbf{17},
022901--022901-8 (2010)
%
\bibitem[\protect\citeauthoryear{Penna et al.}{2010}]{pen10}
R. F. Penna, J. C. McKinney, R. Narayan, A. Tchekhovskoy, R. Shafee,
J. E. McClintock, MNRAS, \textbf{408}, 752-782 (2010)
%
\bibitem[\protect\citeauthoryear{Pessah, Chan \& Psaltis}{2007}]{pes07}
M. E. Pessah, C.-K. Chan, D. Psaltis, ApJ, \textbf{668}, L51-L54 (2007)
%
\bibitem[\protect\citeauthoryear{Poutanen et al.}{2007}]{pou07}
J. Poutanen, G. Lipunova, S. Fabrika, A. G. Butkevich, P. Abolmasov, MNRAS,
\textbf{377}, 1187-1194 (2007)
%
\bibitem[\protect\citeauthoryear{Proga, Stone \& Kallman}{2000}]{pro00}
D. Proga, J. M. Stone, T. R. Kallman, ApJ, \textbf{543}, 686-696 (2000)
%
\bibitem[\protect\citeauthoryear{Quataert, Dorland \& Hammett}{2002}]{qua02}
E. Quataert, W. Dorland, G. W. Hammett, ApJ, \textbf{577}, 524-533 (2002)
%
\bibitem[\protect\citeauthoryear{Quataert \& Gruzinov}{2000}]{qua00}
E. Quataert, A. Gruzinov, ApJ, \textbf{539}, 809-814 (2000)
%
\bibitem[\protect\citeauthoryear{Rees et al.}{1982}]{ree82}
M. J. Rees, M. C. Begelman, R. D. Blandford, E. S. Phinney, Nature,
\textbf{295}, 17-21 (1982)
%
\bibitem[\protect\citeauthoryear{Riquelme  et al.}{2012}]{riq12}
M. A. Riquelme, E. Quataert, P. Sharma, A. Spitkovsky, ApJ,
\textbf{755}, 50, 20 pp. (2012)
%
\bibitem[\protect\citeauthoryear{R\'o\.za\'nska \& Czerny}{2000}]{roz00}
A. R\'o\.za\'nska, B. Czerny, A\&A, 360, 1170-1186 (2000)
%
\bibitem[\protect\citeauthoryear{Sadowski et al.}{2012}]{sad12}
A. Sadowski, R. Narayan, A. Tchekhovskoy, Y. Zhu, MNRAS, in press (2012),
arXiv:1212.5050
%
\bibitem[\protect\citeauthoryear{Sakimoto \& Coroniti}{1981}]{sak81}
P. J. Sakimoto, F. V. Coroniti, ApJ, \textbf{247}, 19-31 (1981)
%
\bibitem[\protect\citeauthoryear{Shakura \& Sunyaev}{1973}]{sha73}
N. I. Shakura, R. A. Sunyaev, A\&A, \textbf{24}, 337-355 (1973)
%
\bibitem[\protect\citeauthoryear{Shakura \& Sunyaev}{1976}]{sha76}
N. I. Shakura, R. A. Sunyaev, MNRAS, \textbf{175}, 613-632 (1976)
%
\bibitem[\protect\citeauthoryear{Shapiro, Lightman \& Eardley}{1976}]{shap76}
S. L. Shapiro, A. P. Lightman, D. M. Eardley, ApJ, \textbf{204}, 187-199
(1976)
%
\bibitem[\protect\citeauthoryear{Sharma, Hammett, \& Quataert}{2003}]{sha03}
P. Sharma, G. W. Hammett, E. Quataert, ApJ, \textbf{596},
1121-1130 (2003)
%
\bibitem[\protect\citeauthoryear{Sharma et al.}{2006}]{sha06}
P. Sharma, G. W. Hammett, E. Quataert, J. M. Stone, ApJ, \textbf{637},
952-967 (2006)
%
\bibitem[\protect\citeauthoryear{Sharma et al.}{2007}]{sha07}
P. Sharma, E. Quataert, G. W. Hammett, J. M. Stone, ApJ, \textbf{667},
714-723 (2007)
%
\bibitem[\protect\citeauthoryear{Shi, Krolik \& Hirose}{2010}]{shi10}
J. Shi, J. H. Krolik, S. Hirose, ApJ, \textbf{708}, 1716-1727 (2010)
%
\bibitem[\protect\citeauthoryear{Shibazaki \& H\=oshi}{1975}]{shi75}
N. Shibazaki, R. H\=oshi, Prog. Theor. Phys., \textbf{54}, 706-718
(1975)
%
\bibitem[\protect\citeauthoryear{Simon, Beckwith \& Armitage}{2012}]{sim12}
J. B. Simon, K. Beckwith, P. J. Armitage, MNRAS, \textbf{422},
2685-2700 (2012)
%
\bibitem[\protect\citeauthoryear{Simon \& Hawley}{2009}]{sim09}
J. B. Simon, J. F. Hawley, ApJ, \textbf{707}, 833-843 (2009)
%
\bibitem[\protect\citeauthoryear{Smak}{1984}]{sma84} J. Smak, PASP,
\textbf{96}, 5-18 (1984)
%
\bibitem[\protect\citeauthoryear{Socrates, Davis \& Blaes}{2004}]{soc04}
A. Socrates, S. W. Davis, O. Blaes, ApJ, \textbf{601}, 405-413 (2004)
%
\bibitem[\protect\citeauthoryear{Socrates}{2010}]{soc10}
A. Socrates, ApJ, \textbf{719}, 784-789 (2010)
%
\bibitem[\protect\citeauthoryear{Sorathia, Reynolds \& Armitage}{2010}]{sor10}
K. A. Sorathia, C. S. Reynolds, P. J. Armitage, ApJ, \textbf{712}, 1241-1247
(2010)
%
\bibitem[\protect\citeauthoryear{Stone et al.}{1996}]{sto96} J. M. Stone,
J. F. Hawley, C. F. Gammie, S. A. Balbus, ApJ,
\textbf{463}, 656-673 (1996)
%
\bibitem[\protect\citeauthoryear{Suzuki \& Inutsuka}{2009}]{suz09} T. K.
Suzuki, S.-I. Inutsuka, ApJ, \textbf{691}, L49-L54 (2009)
%
\bibitem[\protect\citeauthoryear{Svensson \& Zdziarski}{1994}]{sve94}
R. Svensson, A. A. Zdziarski, ApJ, \textbf{436}, 599-606 (1994)
%
\bibitem[\protect\citeauthoryear{Taam \& Lin}{1984}]{taa84}R. E. Taam, D. N.
C. Lin, ApJ, \textbf{287}, 761-768 (1984)
%
\bibitem[\protect\citeauthoryear{Takahashi et al.}{2012}]{tak12}
H. R. Takahashi, K. Ohsuga, Y. Sekiguchi, T. Inoue, K. Tomida, ApJ, in press
(2012), arXiv:1212.4910
%
\bibitem[\protect\citeauthoryear{Turner}{2004}]{tur04}
N. J. Turner, ApJ, \textbf{605}, L45-L48 (2004)
%
\bibitem[\protect\citeauthoryear{Turner et al.}{2005}]{tur05}
N. J. Turner, O. M. Blaes, A. Socrates, M. C. Begelman, S. W. Davis, ApJ,
\textbf{624}, 267-288 (2005)
%
\bibitem[\protect\citeauthoryear{Turner et al.}{2003}]{tur03}
N. J. Turner, J. M. Stone, J. H. Krolik, T. Sano, ApJ,
\textbf{593}, 992-1006 (2003)
%
\bibitem[\protect\citeauthoryear{Wu}{1997}]{wu97}X.-B. Wu, MNRAS, \textbf{292},
113-119 (1997)
%
\bibitem[\protect\citeauthoryear{Wu}{1996}]{wu96}X.-B. Wu, Q.-B. Li, ApJ,
\textbf{469}, 776-783 (1996)
%
\bibitem[\protect\citeauthoryear{Yamasaki}{1997}]{yam97}T. Yamasaki
PASJ, \textbf{49}, 227-233 (1997)
%
\bibitem[\protect\citeauthoryear{Yuan}{2001}]{yua01}F. Yuan,
MNRAS, \textbf{324}, 119-127 (2001)
%
\bibitem[\protect\citeauthoryear{Yuan}{2003}]{yua03}F. Yuan,
ApJ, \textbf{594}, L99-L102 (2003)
\end{thebibliography}
\nocite{*}


\end{document}